\documentstyle[12pt,preprint,aps,floats]{revtex}
\input epsf.tex
\def\DESepsf(#1 width #2){\epsfxsize=#2 \epsfbox{#1} \vspace*{0.07in}}
\def\to{\rightarrow}
\def\sn{\tilde\nu}
\def\mupmum{\mu^+\mu^-}
\def\cnone{\chi^0}
\def\mcnone{m_{\cnone}}

\def\snt{\tilde\nu_\tau}

\def\msnt{m_{\snt}}
\def\dm{\Delta\msnt}

\def\gamsnt{\Gamma_{\snt}}
\def\gev{~{\rm GeV}}
\def\sigsntbar{\overline \sigma_{\snt}}
\def\rp{\not{\hbox{\kern-4pt $R_P$}}}
\def\met{\not{\hbox{\kern-4pt $E_T$}}}
\def\mpt{\not{\hbox{\kern-4pt $p_T$}}}
\def\fbi{ \text{ fb}^{-1}}
\def\anti{\overline}
\def\rts{\sqrt s}
\def\sigrts{\sigma_{\!\rts}}
\tighten

\begin{document}

\preprint{
\noindent
\begin{minipage}[t]{3in}
\begin{flushleft}
November 1997 \\
\end{flushleft}
\end{minipage}
\hfill
\begin{minipage}[t]{3in}
\begin{flushright}
UCD--97--25\\
LBNL--41058\\
UCB--PTH--97/55\\
hep-ph/9711414\\
\vspace*{.7in}
\end{flushright}
\end{minipage}
}

\title{
$\bbox{R}$-Parity Violation and Sneutrino Resonances 
at Muon Colliders}
\author{Jonathan L. Feng$^{a}$, John F. Gunion$^{b}$, and 
Tao Han$^{b,c}$
\vspace*{.2in}
}
\address{${}^{a}$Theoretical Physics Group, Lawrence Berkeley
National Laboratory \\ and Department of Physics, University of
California, Berkeley, CA 94720}
\address{${}^{b}$Davis Institute for High Energy Physics \\
Department of Physics, University of California,
Davis, CA 95616}
\address{${}^{c}$Department of Physics, University of Wisconsin,
Madison, WI 53706
\vspace*{.2in}
}

\maketitle

\begin{abstract}
In supersymmetric models with $R$-parity violating interactions,
sneutrinos may be produced as $s$-channel resonances at $\mu^+ \mu^-$
colliders.  We demonstrate that, for $R$-parity violating couplings as
low as $10^{-4}$, sneutrinos can be discovered and their couplings
measured to high accuracy.  The excellent beam energy resolution of
muon colliders is found to be especially useful for studying such
resonances in certain cases.
\end{abstract}



Low-energy supersymmetry (SUSY) provides an elegant solution to the
gauge hierarchy problem and is a leading candidate for physics beyond
the standard model (SM).  If supersymmetric particles are discovered,
the primary goal of future colliders will be to measure their
properties and thereby determine the SUSY model parameters with high
accuracy.

In many SUSY models, all interactions are assumed to be invariant
under $R$-parity, where $R_P =+1$ $(-1)$ for ordinary SM particles
(superpartners).  If $R$-parity is conserved, all superpartners must
be produced in pairs and the lightest supersymmetric particle (LSP) is
stable.  However, renormalizable gauge-invariant interactions that are
explicitly $R$-parity violating ($\rp$) are also allowed by the
superpotential
\begin{equation}
W = \lambda L L E^c + \lambda' L Q D^c + \lambda'' U^c D^c D^c \ ,
\label{superpotential}
\end{equation}
where the lepton and quark chiral superfields $L = (N, E)$, $E^c$, $Q
= (U, D)$, $U^c$, and $D^c$ contain the standard model fermions $f$
and their scalar partners $\tilde{f}$, and generational indices have
been suppressed.  The first two terms of Eq.~(\ref{superpotential})
violate lepton number $L$, while the last term violates baryon number
$B$.  To avoid proton decay, $L$- and $B$-violating interactions
cannot both be present.

With the couplings of Eq.~(\ref{superpotential}), the scalar partners
of SM fermions may be produced singly at colliders.  In particular,
sneutrinos $\sn$ may be produced as $s$-channel resonances at lepton
colliders~\cite{snu,bgh,efp}.  Such resonance production is unique in
that it probes supersymmetric masses up to $\sqrt{s}$ with large
statistics. As sneutrinos are likely to be among the lighter
superparticles, even a first stage muon collider with $\sqrt{s} = 80 -
250 \text{ GeV}$ will cover much of the typically expected mass range.
Further, the distinctive decays of sneutrinos in $\rp$ SUSY theories
ensure their early discovery at general purpose colliders.  Muon
colliders, with planned luminosities of ${\cal L} \sim 0.1 - 1\fbi
/\text{yr}$ and exceptionally high beam energy resolutions, have been
shown to provide excellent opportunities for studying narrow-width
Higgs boson resonances~\cite{bbgh}.  Here, we demonstrate that, once
the sneutrino mass is approximately known, $\rp$ sneutrino studies
also hold great promise at muon colliders.

We may write the $L$-violating terms of Eq.~(\ref{superpotential}) as
\begin{equation}
\lambda_{ijk} ( N_i E_j E^c_k - E_i N_j E^c_k )
+ \lambda'_{lmn} ( N_l D_m D^c_n - V^*_{pm} E_l U_p D^c_n)\ ,
\label{NDDbasis}
\end{equation}
where $i<j$, all other generational indices are arbitrary, and $V$ is
the CKM matrix.  At muon colliders, sneutrinos $\sn_e$ and
$\sn_{\tau}$ may be produced in the $s$-channel through $\lambda$
couplings.  They can then decay through $\lambda$ ($\lambda'$)
couplings to charged lepton (down-type quark) pairs with widths
$\Gamma (\tilde{\nu}\rightarrow f\anti{f}') = N_{c}
({h^{2}}/{16\pi})m_{\tilde{\nu}}\,,$ where $N_c$ is the color factor
and $h$ is the relevant $\rp$ coupling.  Sneutrinos may also have
$R_P$-conserving decays, such as $\sn \to \nu \cnone$, where the
lightest neutralino $\cnone$ subsequently decays to three SM fermions
through $\rp$ interactions.

\begin{figure}[t]
\centerline{\DESepsf(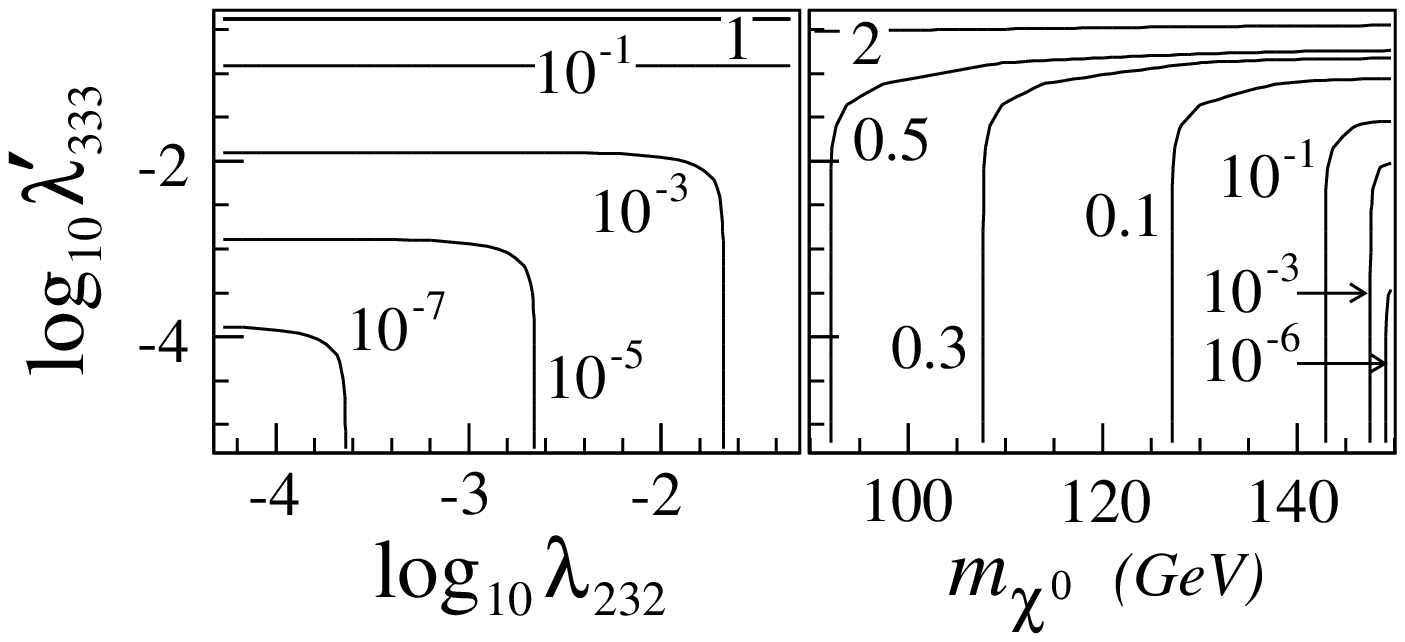 width 0.8 \textwidth) }
\caption{Contours of total decay width $\gamsnt$ in GeV (i) for
$\msnt = 100 \gev$, assuming only $\rp$ decays $\snt \to \mu^+\mu^-,
b\anti{b}$ are open, and (ii) for $\msnt = 150\gev$, assuming that
$\snt \to \nu_{\tau} \cnone$ decays are also allowed, with $\cnone =
\tilde{B}$ and fixed $\lambda_{232}=5\times 10^{-5}$. (In (ii), for
any currently allowed $\lambda_{232}$, $\Gamma(\snt\to\mupmum)\ll
\gamsnt$ unless $\mcnone>145\gev$ and $\lambda'_{333}<0.01$.)}
\label{fig:widths}
\end{figure}

The phenomenology of sneutrino resonances is thus rather complicated
in full generality.  However, in analogy with the Yukawa couplings,
$\rp$ couplings involving higher generational indices are usually
expected to be larger.  We therefore focus on $\sn_\tau$ production
through the coupling $\lambda_{232}$, and, in addition to the decay
$\sn_\tau \to \mu^+ \mu^-$, consider the possibility of $\sn_\tau \to
b \anti{b}$ decays governed by $\lambda'_{333}$.  For simplicity, we
take these two $\rp$ couplings to be real and assume that all other
$\rp$ parameters are negligible.  We will also consider a scenario in
which the $R_P$-conserving decay $\snt \to \nu_{\tau} \cnone$ is
important.  Fig.~\ref{fig:widths} shows representative decay widths
for the three modes.

The cross section for resonant $\snt$ production is
\begin{equation}
\sigma_{\snt}(\sqrt{s}) = {8\pi\Gamma(\snt\to\mu^+\mu^-)
\Gamma(\snt\to X)\over (s-\msnt^2)^2 +\msnt^2\gamsnt^2} \ ,
\label{resonancexsec}
\end{equation}
where a factor of 2 has been explicitly included to account for both
$\snt$ and $\snt^*$ exchange, $X$ denotes a generic final state from
$\snt$ decay, and $\gamsnt$ is the total sneutrino decay width.  The
effective cross section $\sigsntbar$ is obtained by convoluting
$\sigma_{\snt}(\rts)$ with the collider's $\rts$ distribution.
Neglecting (for purposes of discussion) bremsstrahlung and
beamstrahlung, this distribution is well-approximated by a Gaussian
distribution with rms width
\begin{equation}
\sigrts = (7~{\rm MeV})\left({R\over 0.01\%}\right)\left({\sqrt
s\over {\rm 100\ GeV}}\right) \, ,
\label{resolution}
\end{equation}
where the beam energy resolution factor $R$ is typically in the range
(0.003 - 0.1)\%.  In two extreme limits, $\sigsntbar$ can be expressed
in terms of branching fractions $B$ as
\begin{equation}
\sigsntbar (\msnt)
 = \left\{ \begin{array}{l}
 \! \frac{\protect\sqrt{8\pi^3}}{\msnt^2} \frac{\gamsnt}{\sigrts}
B(\mu^+ \mu^- ) B(X) \, , \, \gamsnt \ll \sigrts \\
\frac{8\pi}{\msnt^2} B(\mu^+ \mu^- ) B(X) \, , \, 
\gamsnt \gg \sigrts \ . \end{array} \right.
\label{widthxsec}
\end{equation}
If only highly suppressed $\rp$ decays are present, $\sigsntbar \propto
\gamsnt/\sigrts$.  The small values of $\sigrts$ possible at a muon
collider thus provide an important advantage for probing small $\rp$
couplings. At a muon collider, the effects of bremsstrahlung are small
(but are included in our numerical results); beamstrahlung is
negligible.

There are a wide variety of low-energy constraints on $\lambda_{232}$
and $\lambda'_{333}$ \cite{subpap}; the most stringent of these are
collected in Table~\ref{table:t1}. Bounds on these individual
couplings may be found in the literature.  Their product is most
stringently bounded by $B$ decays~\cite{efp,jkl}.  In the $NDD^c$
diagonal basis of Eq.~(\ref{NDDbasis}), the product is most tightly
constrained by applying the bound $B(B^- \to \mu^- \anti{\nu}) <
2.1\times 10^{-5}$~\cite{pdg} to the operator $-\frac{\lambda_{232}
\lambda'_{333}} {m_{\tilde{\tau}_L}^2} V_{pb} (\overline{\mu_R}
\nu_L)(\overline{u_L^p} b_R)$.  In the alternative $EUD^c$ diagonal 
basis, the $\rp$ interactions $\lambda'^U_{lmn} \left( V_{mq} N_l D_q
D^c_n - E_l U_m D^c_n \right)$ generate the operator $-
\frac{\lambda_{232}\lambda'^U_{333}} {m_{\tilde{\nu}_\tau}^2} V^*_{tq}
(\overline{\mu_R} \mu_L)(\overline{d_L^q} b_R)$, which is stringently
bounded by $B(B_s \to \mu^+ \mu^-) < 7.7 \times
10^{-7}$\cite{bsbound}.  {}From Table~\ref{table:t1} we see that
bounds on the product are highly basis-dependent.  We shall not assume
that $\lambda'$ is diagonal in any particular basis, but only that
$\lambda'_{333}$ is dominant. For typical superpartner masses,
appropriate limits are then $\lambda_{232} \alt 0.06$, $\lambda'_{333}
\alt 1$, and $\lambda_{232} \lambda'_{333} \alt 0.001$.
\begin{table}
\caption{Upper bounds on the couplings $\lambda_{232}$ and
$\lambda'^{(U)}_{333}$.}
\begin{tabular}{lll}
Coupling & Upper Bound & Process \\
\hline
$\lambda_{232}$ & 0.06 $[m_{\tilde{\mu}_R}/100 \text{ GeV}]$ &
$\frac{\Gamma(\tau\to e \nu \anti{\nu})}{\Gamma(\tau\to \mu \nu
\anti{\nu})}$\cite{bgh,dreiner,pdg}
\\
$\lambda'_{333}$ & $0.6 - 1.3$ ($2\sigma$) &
$R_{\tau}$ ($m_{\tilde{q}} = 0.3-1$ TeV)~\cite{bes}
\\
$\lambda_{232}\lambda'_{333}$ & $0.089 \left[{m_{\tilde{\tau}_L}/ 100
\text{ GeV}}\right]^2$
& $B^- \to \mu \anti{\nu}$
\\
$\lambda'^U_{333}$ & $1.0 \left[m_{\tilde{\nu}_{\tau}}/ 100 \text{
GeV}\right]^{1/2}$
&
$B^0$--$\anti{B}^0$ ($2m_{\tilde{\nu}_{\tau}} = 
m_{\tilde{b}_R}$)~\cite{ag}
\\
$\lambda_{232}\lambda'^U_{333}$ & 0.0012
$\left[{m_{\tilde{\nu}_{\tau}}/ 100 \text{ GeV}}\right]^2$
&
$B_s \to \mu^+ \mu^-$
\\
\end{tabular}
\label{table:t1}
\end{table}

The signals for $\snt$ production depend on the $\snt$ decay
patterns. We consider two well-motivated scenarios. In the first,
$\msnt < \mcnone$, and $\snt$ decays only through $\rp$
operators.\footnote{If, as expected, right-handed sleptons are lighter
than sneutrinos, $\sn \to \nu l_R \tilde{l}_R$ will be kinematically
allowed.  However, such decays are competitive with $\rp$ decays only
if (i) the $\rp$ decays have the very small widths found for $\rp$
couplings close to or below the discovery limit and (ii) $\tilde{B}$
and $\tilde{l}_R$ are very light such that the $R_P$-conserving decay
is nearly maximal.  We therefore neglect decays to $\tilde{l}_R$ in
this analysis.}  Neglecting $\rp$ couplings other than $\lambda_{232}$
and $\lambda'_{333}$, the signal is $\mu^+ \mu^-$ or $b\anti{b}$ pairs
in the final state.  For concreteness, we consider $\msnt = 100\gev$.

The dominant backgrounds are Bhabha scattering and $\mu^+\mu^- \to
\gamma^*, Z^* \to \mu^+\mu^-, b\anti{b}$.  To reduce these, we apply 
the following cuts: for the $\mu^+\mu^-$ ($b\anti b$) channel, we
require $60^{\circ} <\theta < 120^{\circ}$ ($10^{\circ} < \theta <
170^{\circ}$) for each muon ($b$ quark). The stronger $\theta$ cuts in
the $\mupmum$ channel are needed to remove the forward-peaked Bhabha
scattering.  We also require $|m_{f\anti{f}} - \msnt | < 7.5\gev$ in
both channels to reduce background from radiative returns to the $Z$.

\begin{figure}[t]
\centerline{\DESepsf(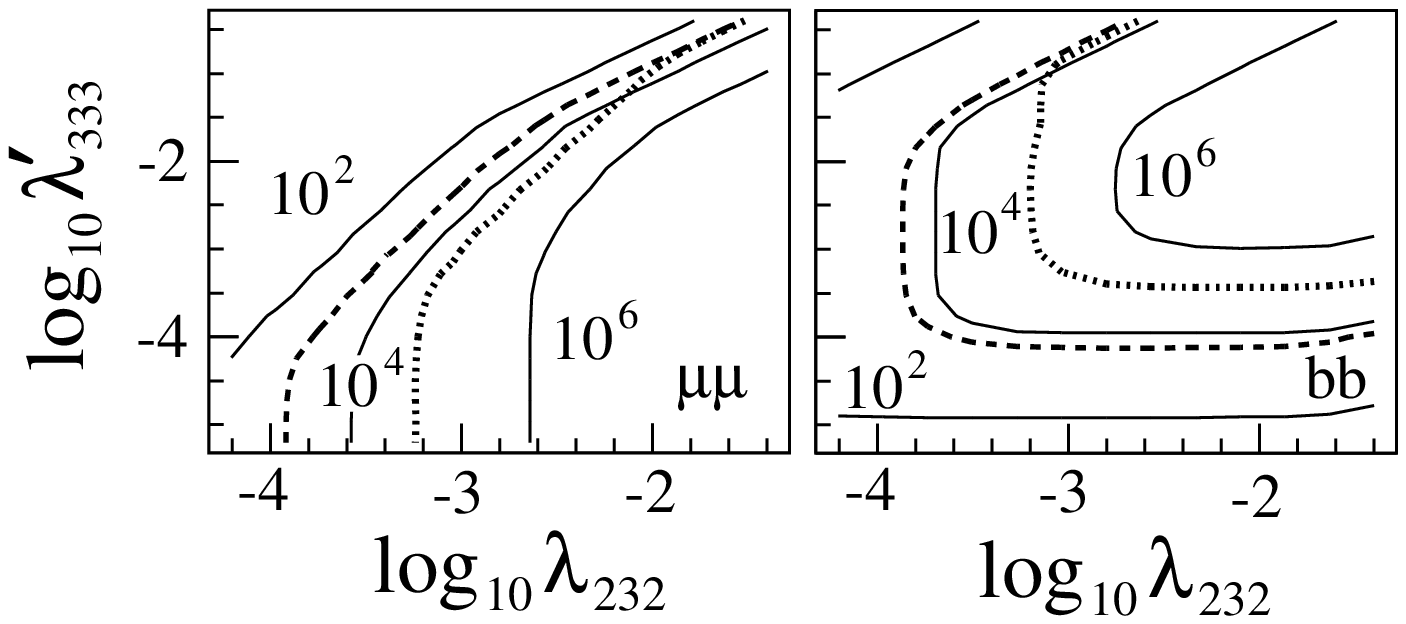 width 0.8 \textwidth) }
\caption{Contours for (i) $\sigma(\mu^+ \mu^- \to
\snt \to \mu^+ \mu^- )$ and (ii) $\sigma(\mu^+ \mu^- \to
\snt \to b\anti{b})$ (solid) in fb after cuts for the $\msnt 
< m_{\cnone}$ scenario, with $\protect\sqrt{s} = \msnt= 100\gev$ and
$R=0.003\%$.  The dashed and dotted contours give the optimistic and
pessimistic/scan $3\sigma$ discovery boundaries, respectively, for
total integrated luminosity $L = 0.1\fbi$.  (See discussion in text.)}
\label{rates1}
\end{figure}

After the cuts above and including beam energy spread and
bremsstrahlung, the background cross sections at $\sqrt{s} = 100\gev$
are $\sigma(\mupmum)=3.5\times 10^4~{\rm fb}$ and $\sigma(b\anti
b)=2.0\times 10^5~{\rm fb}$.  Given the two options of $({\cal L},R)=
(0.1\fbi/\text{yr}, 0.003\%)$ and $(1\fbi/\text{yr}, 0.1\%)$, we find
that the former maximizes $S/\sqrt B$ when $\gamsnt$ is small. In this
scenario, $\gamsnt$ is unknown {\em a priori}, but since a very small
$\gamsnt$ is quite possible (see Fig.~\ref{fig:widths}), we employ
this option.  With this choice, signal cross sections after cuts are
given by the solid contours in Fig.~\ref{rates1}. We see that the
cross sections may be extremely large ($>1$ nb) in some regions of the
allowed parameter space.

In Fig.~\ref{rates1} we also give sneutrino resonance discovery
contours for two extreme possibilities. In the most optimistic case,
the sneutrino mass is exactly known and the total luminosity is
applied at the sneutrino resonance peak. The corresponding
``optimistic'' $3\sigma$ discovery contours are given by dashed
lines. (In calculating $S/\sqrt{B}$ for the $b\anti{b}$ mode here and
below, we include a 75\% efficiency for tagging at least one $b$
quark.) More realistically, the sneutrino mass will be known only
approximately from other colliders with some uncertainty $\pm
\frac{1}{2} \dm$; we assume $\dm = 100\text{ MeV}$ using the fully
reconstructable $\rp$ decays.  To ensure significant overlap with the
$\snt$ resonance, the luminosity should be distributed uniformly
within this interval into $N=(\dm/\Delta\rts)+1$ scan points, where
$\Delta \rts=\text{max}[2\sigrts,\gamsnt]$.  As noted above, $\gamsnt$
is unknown beforehand in this scenario; to allow for the possibility
of very narrow widths, we choose $\Delta \rts= 2\sigrts$.  We
conservatively assume that all $N$ points must be scanned before an
observable signal becomes apparent.\footnote{To maximize the
probability of discovering a broad resonance early, the scan should be
optimized by ordering the scan points such that, at any given time,
the completed scan points are roughly uniformly distributed in the
$\dm$ interval.}  If $\gamsnt < \sigrts$, the worst possible case is
when $\msnt$ lies midway between two scan points.  For discovery, we
demand a $3\sigma$ excess in the combined data of these two points.
If the width is large, the most difficult possibility is that $\msnt$
lies at one end of the mass interval, so luminosity is collected in
only one half of the peak. We demand a 3$\sigma$ excess in the
combined data collected at scan points lying within $\gamsnt/2$ of
$\msnt$. The resulting ``pessimistic/scan'' $\snt$ discovery
boundaries are the dotted contours of Fig.~\ref{rates1}. The actual
discovery limit should lie between the dashed and dotted contours.  We
see that $\snt$ resonance observation is possible for $\rp$ couplings
as low as $10^{-3} - 10^{-4}$.

We now consider a second scenario in which $m_{\sn_\tau} >
m_{\cnone}$.  In addition to the $\rp$ decays of the previous
scenario, decays $\snt\to\nu_{\tau} \cnone$ are therefore allowed and
typically dominate, with $\cnone$ then decaying to $\nu_{\tau}\mu\mu$
or $\nu_{\mu} \mu \tau$ through the $\lambda_{232}$ coupling, or
$\nu_{\tau}b\anti{b}$ through the $\lambda'_{333}$ coupling.  We
assume that no other $\snt$ decays are significant, and $\cnone \to
l_R \tilde{l}_R$ is closed.  (Note that $\snt \to \nu_{\tau} l_R
\tilde{l}_R$, even if kinematically allowed, is again negligible, as 
it is typically highly suppressed relative to $\snt\to\nu_{\tau}
\cnone$.) The final signals are then $\mupmum + \met$, 
$\mu^{\pm}\tau^{\mp} + \met$, and $b\anti{b} + \met$. For this
scenario, we consider masses $\msnt = 150\gev$ and $\mcnone =
100\gev$.

\begin{figure}[t]
\centerline{\DESepsf(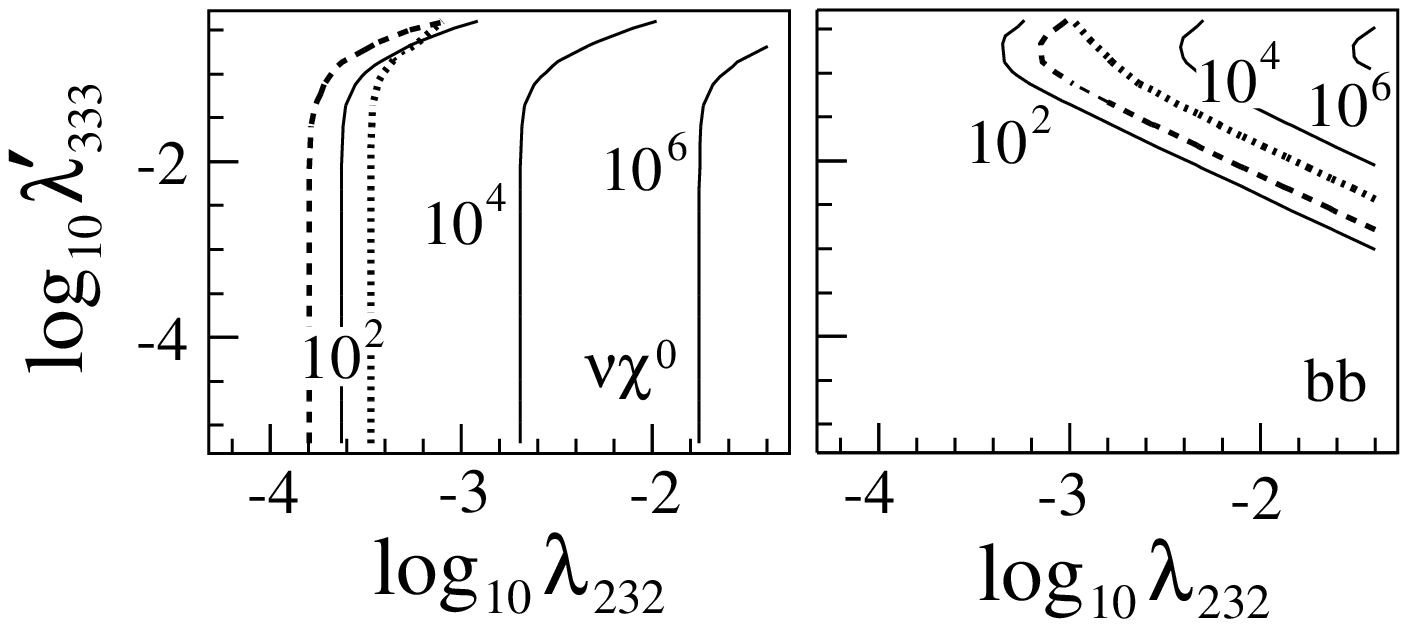 width 0.8 \textwidth) }
\caption{Contours for (i) $\sigma(\mu^+ \mu^- \to
\snt \to \nu\cnone)$ (no cuts) and (ii) $\sigma(\mu^+ \mu^- \to
\snt \to b\anti{b})$ (after cuts) in fb assuming $\msnt = 150\gev$, 
$\mcnone = 100\gev$, and $\cnone= \tilde{B}$.  The optimistic (dashed)
and pessimistic/scan (dotted) discovery contours assume $L=1\fbi$ and
$R=0.1\%$. (See text.)}
\label{rates2}
\end{figure}

The signal cross sections for the $\nu\cnone$ channel (without cuts)
and the direct $\rp$ $b\anti{b}$ channel (after cuts as in
Fig.~\ref{rates1}) are plotted in Fig.~\ref{rates2}.  (The cross
section for the direct $\rp$ $\mupmum$ channel is negligible unless
$\lambda_{232}$ is close to its current upper bound.)  The leading
backgrounds to the $\nu\cnone$ channels are from $WW^{(*)}$ and
$ZZ^{(*)}$. To reduce these, we require $\met > 25 \text{ GeV}$, that
the visible final state fermions have $p_T > 25\gev$ and $60^{\circ} < \theta
< 120^{\circ}$ for the lepton modes ($40^{\circ} < \theta <
140^{\circ}$ for the $b\anti{b}\met$ mode), and that the invariant
mass of the two visible fermions be $ > 50\gev$.  With these cuts,
the total combined background in the $\nu\cnone$ channels is $\sim 1$
fb.

In Fig.~\ref{rates2}, we also give 3$\sigma$ discovery contours for
both optimistic and pessimistic/scan cases as before. In the case of
the $\nu\cnone$ mode, we sum over all final channels and employ
Poisson statistics. The kinematic distributions for the $\met$ modes
depend on the masses of the superpartners entering virtually in the
$\cnone$ decay. As an example, we have taken all sleptons degenerate
with $\snt$ and $m_{\tilde{b}_{L,R}} = 200\gev$.  With these masses,
the overall signal efficiency is 10\% for both the $\mu^+\mu^- + \met$
and $b\anti{b} + \met$ channels (where, in the latter case, the
$b$-tagging efficiency has been included as before); we assume a
similar efficiency for the $\mu^{\pm} \tau^{\mp} + \met$ channel.

The scan requires discussion. In this scenario, the dominant decay
$\snt\to\nu_{\tau} \cnone$ results in missing energy, so direct
reconstruction of $\msnt$ is impossible; the $\snt$ mass measurement
at other colliders must then rely on kinematic endpoints, yielding
typically $\dm\sim 2\gev$.  (More accurate mass measurements are
possible if $\rp$ $b\anti b$ decays are also observed, but we will not
assume this here.)  However, in contrast to the previous scenario,
$\gamsnt$ is large and can be computed based on measurements of the
$\cnone$ mass and composition [and $r_b\equiv B(\snt\to b\anti
b)/B(\snt\to \nu_{\tau} \cnone)$ if significant] at other colliders.
For our choice of parameters, $\gamsnt = 0.36\gev(1+r_b)$, so we scan
with $\Delta \rts=\gamsnt$ and $R=0.1\%$ (which is adequate and allows
maximal ${\cal L}$).

{}From Fig.~\ref{rates2}, we see that the nearly background-free
$\nu\cnone$ mode makes possible a dramatic improvement in discovery
reach compared to the $\msnt<\mcnone$ scenario. The $\snt$ resonance
may be discovered for $\lambda_{232} \agt 10^{-4}$, irrespective of
the value of $\lambda'_{333}$.  Note that the discovery region in the
direct $\rp$ $b\anti{b}$ channel is partially excluded by current
bounds on $\lambda_{232}\lambda'_{333}$.

\begin{figure}[t]
\centerline{\DESepsf(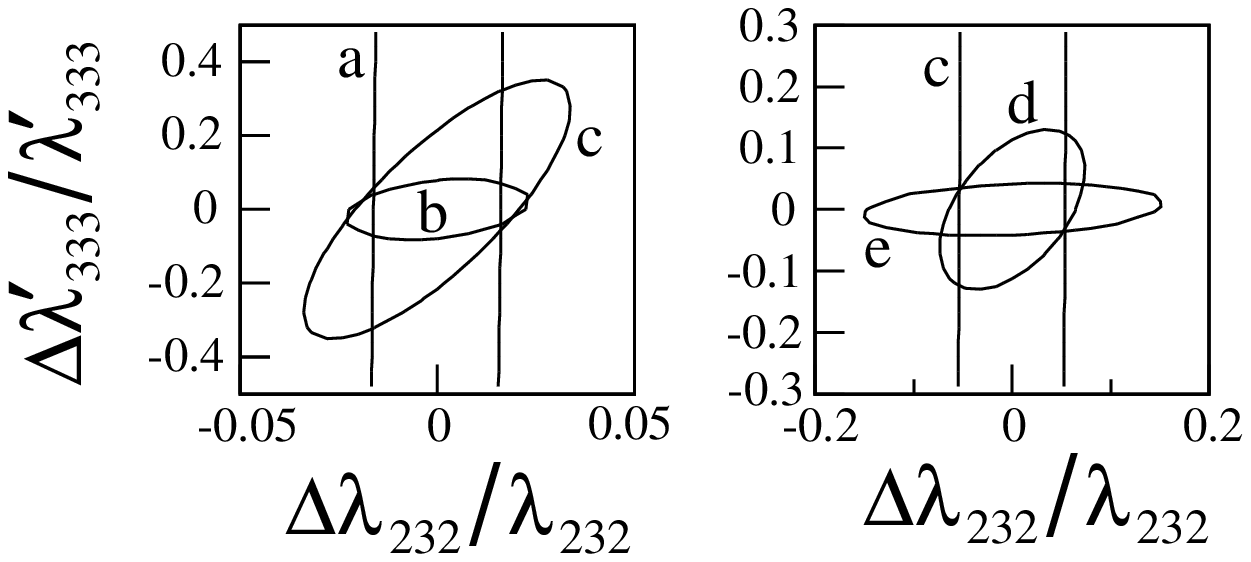 width 0.8 \textwidth) }
\caption{$\chi^2 = 1$ contours in the $(\Delta\lambda_{232} 
/ \lambda_{232}, \Delta\lambda'_{333} / \lambda'_{333})$ plane for (i)
the $\msnt = 100 \gev < \mcnone$ scenario, assuming $L=0.3\fbi$,
$R=0.003\%$ and (ii) the $\msnt = 150\gev$ $>$ $\mcnone = 100\gev$
scenario, assuming $L=3\fbi$, $R=0.1\%$.  Contours are for
$\lambda_{232}=5\times 10^{-4}$ and $\lambda'_{333}=$: (a) $10^{-5}$;
(b) $5\times 10^{-4}$; (c) $10^{-2}$; (d) $10^{-1}$; (e) $0.3$.}
\label{signif}
\end{figure}

Once we have found the sneutrino resonance via the scan described, the
crucial goal will be to precisely measure the relevant $\rp$
couplings. In the $\msnt<\mcnone$ scenario, the discovery scan gives a
precise determination of $\msnt$ (and, if $\gamsnt>2\sigrts$, a rough
determination of $\gamsnt$). We then envision accumulating $L=0.1\fbi$
($R=0.003\%$) at each of the three points $\rts=\msnt$, $\msnt\pm
\Delta\rts/2$, where $\Delta \rts=\text{max}[2\sigrts,\gamsnt]$.  
The off-resonance points ensure good sensitivity to $\gamsnt$.  This
is especially crucial when $\gamsnt>\sigrts$, as in this case a single
measurement of $\sigsntbar$ at $\rts=\msnt$ determines
$B(\snt\to\mupmum)$ but not $\Gamma(\snt\to\mupmum)$; see
Eq.~(\ref{widthxsec}).  In the $\msnt>\mcnone$ scenario, we noted that
$\gamsnt$ can be computed with good precision from observations at
other colliders; we assume a $\pm 5\%$ error for $\gamsnt$. We would
then run only at $\rts\simeq \msnt$ and accumulate $L=3\fbi$
($R=0.1\%$). In Fig.~\ref{signif}, the resulting $\chi^2 = 1$ error
contours are plotted for each of the two scenarios. We find $1\sigma$
errors of $\Delta\lambda_{232}/\lambda_{232}\sim 2-15\%$ for the
representative value of $\lambda_{232}=5\times 10^{-4}$, which is not
very far inside the discovery regions; $\Delta\lambda'_{333} /
\lambda'_{333}\sim 10-30\%$ is achieved if $\lambda'_{333}$ is not too
small.  For small $\lambda'_{333}$ in the $\snt\to\nu_{\tau} \cnone$
scenario, $\Delta\lambda_{232}/\lambda_{232}\sim 3, 6, 10, 38, 110\%$
for $\lambda_{232}=10^{-3},5\times10^{-4},3\times 10^{-4},10^{-4},
5\times10^{-5}$, respectively.  Note that for small $\rp$ couplings,
absolute measurements through other processes and at other colliders
are extremely difficult, as they typically require that $\rp$ effects
be competitive with a calculable $R_P$-conserving process. For
example, $\rp$ neutralino branching ratios constrain only ratios of
$\rp$ couplings.

As a final remark, we note that $\rp$ interactions can split the
complex scalar $\snt$ into a real CP-even and a real CP-odd mass
eigenstate.  This splitting is generated both at tree-level (from
sneutrino-Higgs mixing) and radiatively, and both contributions depend
on many SUSY parameters.  However, such $\rp$ terms also generate
neutrino masses, and it is generally true that the sneutrino
splittings generated are ${\cal O}(m_{\nu})$~\cite{g}.  Given the
current bound $\nu_{\tau} < 18.2 \text{ MeV}$\cite{nutaumass}, we see
that $\tau$ sneutrino splittings may be as large as ${\cal O} (10
\text{ MeV})$. A muon collider with $R=0.003\%$ would be unique in its
ability to resolve splittings of the resonance peak at the MeV level
or better.  The discovery of a non-zero $\nu_{\tau}$ mass would be a
significant motivation for exploring possible sneutrino splittings to
high precision.

In summary, we have demonstrated that a muon collider is an excellent
tool for discovering sneutrino resonances and measuring their
$R$-parity violating couplings.  In addition, such a collider is
unique in its ability to resolve the splitting between the CP-even and
CP-odd sneutrino components when this splitting is as small as
expected given the current bounds on neutrino masses. Further details
will appear in a subsequent paper~\cite{subpap}.

\acknowledgements

The authors acknowledge important discussions with R.~Hempfling and,
especially, V. Barger, and thank the Aspen Center for Physics for
hospitality. This work was supported in part by the Director, Office
of Energy Research, Office of High Energy and Nuclear Physics,
Division of High Energy Physics of the DOE under Contracts
DE--AC03--76SF00098 and DE-FG03-91ER40674, by the NSF under grant
PHY--95--14797, and by the U.C. Davis Institute for High Energy
Physics.

\end{document}